\documentstyle[prb,aps,preprint]{revtex} 
\begin{document}
\draft
\title{Calculation of the singlet-triplet gap of the\\
antiferromagnetic Heisenberg model on the ladder}
\author{M. Azzouz, L. Chen, and S. Moukouri}
\address{Centre de Recherche en Physique du Solide
et D\'epartement de Physique\\ Universit\'e de Sherbrooke, Sherbrooke
Qu\'ebec,
 J1K 2R1 Canada}
\date{\today}
\maketitle
\begin{abstract}
The ground state energy and the
singlet-triplet energy gap of the
antiferromagnetic
Heisenberg model on a ladder is investigated using a mean field
theory and the density matrix renormalization group. Spin wave theory
shows that the corrections to the local magnetization are infinite.
This indicates that no long range order occurs in this system.
A flux-phase state is used to calculate the energy gap as a function
of the transverse coupling, $J_\perp$, in the ladder. It is
found that the gap is linear
in $J_\perp$ for $J_\perp\gg 1$ and goes to zero for $J_\perp\to 0$.
The mean field theory agrees well with the numerical results.
\end{abstract}
\section{introduction}
The antiferromagnetic (AF) Heisenberg {\it ladders} (two coupled spin chains)
are relevant to the
understanding of how the physics evolves from the purely one-dimensional
(1D) systems to two-dimensional (2D). Also,
the $S=1/2$ Heisenberg model on the ladder can model the magnetic
properties of systems such as the vanadyl pyrophosphate \cite{Johnston},
(VO)$_2$P$_2$O$_7$. The calculation of the thermodynamic
properties of this material will be addressed in a forth coming work
\cite{Butaud}. The relevance of the
transverse coupling $J_\perp$ for ground state properties is examined
in the present work.

	From the theoretical point of view, two main reasons are behind
the increasing interest in the Heisenberg ladder. The Haldane conjecture
\cite{Haldane} that the energy gap of the
elementary excitations of chains depends on whether the spin is integer or
half-integer.
It is well known \cite{Cloiseaux} that the spin one-half
1D Heisenberg model is gapless.
However it is not clear how the gap behaves when a transverse coupling,
$J_\perp$ is turned on between the two chains of the Heisenberg ladder.
In Ref. 5 the authors reported that a finite transverse coupling
${J_\perp}_c\approx 0.4$ is
required to get a finite gap. Latter on, Barnes $et$ $al$ \cite{Barnes}
concluded
that the critical value vanishes ${J_\perp}_c=0$. The second reason is the
discovery of high critical temperature superconductors (HT$_c$S). The
interest for these systems is due to the belief that the 2D Heisenberg
model describes the AF interactions in the undoped copper-oxygen planes
of HT$_c$S. The ladder problem can help understand the crossover
from 1D to 2D systems (this question is related to
the stability of the Luttinger liquid).

	Strong and Millis \cite{Strong} have also recently used this type
of model to investigate the competition between magnetic ordering and
the Kondo effect in heavy fermion systems. Since the work of Doniach
\cite{Doniach}, it is widely believed that spin excitations of heavy
fermions can be modeled by such a model. The major weakness is
that the Kondo effect is lost; only spin competition effect is shown.

	The problem of a plane of coupled spin $1/2$ chains was studied by
Azzouz \cite{Azzouz} who found within a mean field approximation that
once $J_\perp$ is non-zero long range AF order appears in the system.
The gap remains zero because of the broken symmetry due to long range order.
For the ladder, no broken symmetry is expected to occur because of the
1D nature of the system. Indeed,
the finite size in the transverse direction will bring different physics
than the two dimensional Heisenberg model. We believe that the gap starts
to be non-zero for any finite $J_\perp$ due to the finite size
in the transverse direction \cite{Parola}. In the
limit of large $J_\perp$ the system is equivalent to a weakly coupled singlets
and the gap is given in leading order by the singlet-triplet energy
separation. The first term of the
gap is linear in $J_\perp$. When $J_\perp$ becomes of the same order as the
parallel coupling the situation becomes more difficult to analyze.

	In this paper, the ladder problem is investigated using a
mean-field approach and exact diagonalization.
The energy gap is found to
be non-zero for any finite $J_\perp$. Comparison with the exact
diagonalization which is based on the density matrix renormalization group
(DMRG) \cite{White}
is reported. The agreement between these approaches is very good.
\section{Mean field treatment}
The Heisenberg model on the ladder is denoted as follows
\begin{equation}
H=J_\perp\sum_{\langle i,j\rangle_\perp}{\bf S}_i\cdot{\bf S}_j
+J\sum_{\langle i,j\rangle_\parallel}{\bf S}_i\cdot{\bf S}_j
\end{equation}
where the sums run over first nearest neighbors
$\langle i,j\rangle_\parallel$ along the chains and
$\langle i,j\rangle_\perp$ perpendicular to the chains. $J$ and $J_\perp$
are AF coupling constants. In the following $J$ is set equal to unity and
periodic boundary conditions are imposed. Simple
limits of this model can be analyzed. The first obvious one is realized
for $J_\perp\gg 1$ as mentioned in the Introduction. In this case,
one gets weakly coupled (by $J=1$) singlets
and the first excited state has an energy gap behaving as
$J_\perp$ in leading order. This excitation is obtained when the
state of a single pair of spins changes from singlet to triplet. The
second limit, less obvious, is $J_\perp=0$. The two chains can be treated
separately. It is known from the exact results of des Cloiseaux
\cite{Cloiseaux} that the 1D Heisenberg model has an energy spectrum of
the form
\begin{equation}
\epsilon(k)={\pi\over2}|\sin k|
\label{spectre1}
\end{equation}
which shows a zero gap. The intermediate regime ($J_\perp\sim 1$) is
quite interesting and is the most complicated one. The dependence of the
energy gap, hereafter denoted
$E_g(J_\perp)$, on the transverse coupling is investigated here using the same
mean-field theory as in Ref. 7.
\subsection{Review of the flux-phase and N\'eel-flux-phase states}
The 2D generalization of Wigner-Jordan transformation of Ref.
7 can be easily
implemented in the case of the ladder. One gets, following the notation
of Fig. 1
\begin{equation}
S^-_{i,1}=c_{i,1}\exp\biggl[i\pi \sum_{\ell=0}^{i-1}\biggl(n_{\ell,1}
+n_{\ell,2}\biggr)\biggr]
\end{equation}
for the chain $1$ and
\begin{equation}
S^-_{i,2}=c_{i,2}\exp\biggl[i\pi \biggl(\sum_{\ell=0}^{i}n_{\ell,1}
+\sum_{\ell=0}^{i-1}n_{\ell,2}\biggl)\biggr]
\end{equation}
for the chain $2$. The indices $i$ run along the chains. The Hamiltonian
is now written in this fermion representation. One finds the following
spinless interacting fermion Hamiltonian :
\begin{eqnarray}
H=&&{-J\over2}\sum_{i,\delta}\biggl[c_{i,1}e^{-i\Phi_{i,i+\delta}(1)}
c^{\dag}_{i+\delta,1} +
c_{i,2}e^{-i\Phi_{i,i+\delta}(2)}c^{\dag}_{i+\delta,2}
\biggr]
+{-J_{\perp}\over2}\sum_{i}c_{i,1}c^{\dag}_{i,2}\cr
&&+J\sum_{i,j=1;2,\delta}(n_{i,j}-1/2)(n_{i+\delta,j}-1/2)+J_{\perp}\sum_i
(n_{i,1}-1/2)(n_{i,2}-1/2)
\label{hamiltonien}
\end{eqnarray}
The phases $\Phi$ are as follows
\begin{equation}
\begin{array}{ll}
\Phi_{i,i+1}(1)&=\pi n_{i,2}\\
\Phi_{i,i-1}(1)&=-\pi n_{i-1,2}\\
\Phi_{i,i+1}(2)&=\pi n_{i+1,1}\\
\Phi_{i,i-1}(2)&=-\pi n_{i,1}
\end{array}
\end{equation}
and $\delta$ refers to the first nearest neighbors of a given site.
The mean field solutions studied here are the flux-phase
\cite{Affleck},
with zero magnetization and the N\'eel-flux-phase with finite magnetization
$m\equiv 2\langle n_{i,j}\rangle-1$. The flux due to the XY term (the
first and second terms in Eq.\ (\ref{hamiltonien})) of the
Hamiltonian is taken to be $\pi$ per plaquette on average.
For the Ising term (the last two terms of Eq.\ (\ref{hamiltonien})),
one chooses
$\langle c_{i,j}c_{i+\delta,j}^{\dag}\rangle=
|\langle c_{i,j}c_{i+\delta,j}^{\dag}\rangle|e^{-i\theta_{i,i+\delta}(j)}$ and
$\langle c_{i,j}c_{i,j+\delta}^{\dag}\rangle=
|\langle c_{i,j}c_{i,j+\delta}^{\dag}\rangle|
e^{-i\theta'_{j,j+\delta}(i)}$ ($j=1,2$).
In the following, we set $|\langle c_{i,j}c_{i+\delta,j}^{\dag}\rangle|=Q$ and
$|\langle c_{i,j}c_{i,j+\delta}^{\dag}\rangle|=P$.
The sum over $\theta$'s around one plaquette is also taken to be $\pi$ on
average. The bipartite character and the different phases on each link of
the system are summarized in Fig. 2. Despite the fact that a finite
magnetization in a system like the ladder is not possible because of its
smallness, we will discuss
the N\'eel-flux-phase and compare its physical $meaning$ with the more
physical flux phase state.
\subsection{Results and discussion}
	The mean field Hamiltonian is written as follows
\begin{eqnarray}
H=&&{J\over2}\sum_{i}\biggl(c^{\dag}_{i,1}e^{-i\pi}c_{i+1,1} +
c^{\dag}_{i,1}c_{i-1,1}+
c^{\dag}_{i,2}c_{i+1,2}+c^{\dag}_{i,2}e^{i\pi}c_{i-1,2}
\biggr)\cr
&&+{J_{\perp}\over2}\sum_{i}\biggl(c^{\dag}_{i,1}c_{i,2}
+c^{\dag}_{i,2}c_{i,1}\biggr)\cr
&&+{J\over2}\sum_{i,j=1,2,\delta}\biggl(mn_{i,j}-mn_{i+\delta,j}+m^2/2\biggr)
+{J_{\perp}\over2}\sum_i\biggl(mn_{i,1}-mn_{i,2}+m^2/2\biggr)\cr
&&+{J}\sum_{i,j=1,2}\biggl(Qc_{i,1}^{\dag}e^{-i\pi}c_{i+1,1}
+Qc^{\dag}_{i,1}c_{i-1,1}
+Qc_{i,2}^{\dag}c_{i+1,2}+Qc^{\dag}_{i,2}e^{i\pi}c_{i-1,2}+Q^2\biggr)\cr
&&+{J_\perp}\sum_{i}\biggl(Pc_{i,1}^{\dag}c_{i,2}+
Pc_{i,2}^{\dag}c_{i,1}+P^2\biggr)
\label{moyen1}
\end{eqnarray}
where a bipartite lattice due to AF correlations is used, Fig. 2 (the local
magnetization is staggered: $m=m_A=-m_B$ where $A$ and $B$ are two adjacent
sites). We get
\begin{equation}
H=\sum_{\bf k}E_{\pm}({\bf k})\alpha_{\bf k,\pm}^{\dag}\alpha_{\bf k,\pm}
\end{equation}
in k-space where the fermionic operator $\alpha_{\bf k}$ is obtained from
$c_{\bf k}$ through the diagonalization of the Hamiltonian (\ref{moyen1}).
The dispersion relation is given by
\begin{eqnarray}
E_{\pm}({\bf k})=\pm\biggl(m^2(1+J_\perp/2)^2+(1+2Q)^2\sin^2k_x+
(1+2P)^2{(J_\perp/2)}^2\cos^2k_y\biggr)^{1/2}
\label{dispersion}
\end{eqnarray}
where $J_\perp$ is divided by a factor $2$ since
the periodic boundary conditions
used in the transverse direction count $J_\perp$ twice.
The minimization of the free energy with respect to $m$, $Q$ and $P$ gives
a set of three self-consistent equations which become
\begin{equation}
\begin{array}{ll}
&m=\int {({d^2k/2\pi})} m{{(1+J_\perp/2)}/{E_+({\bf k})}},\\
&Q=\int {({d^2k}/{2\pi})} {{(1+2Q)\sin^2k_x}/{E_+({\bf k})}},\\
&P=\int {({d^2k}/{2\pi})} {{(1+2P)(J_\perp/2)\cos^2k_y}
/{E_+({\bf k})}}
\label{moyen}
\end{array}
\end{equation}
at zero temperature. The integration is over the first Brillouin zone.
By definition, we write
$\int {{d^2k}/{2\pi}}\equiv \int {({dk_x}/{2\pi})}(1/2)\sum_{k_y}$ where
$k_y$ can take two values: $0$ or $\pi$.
One easily notes that $m=0$ is a solution. An interesting feature
shown by such a solution is that when $J_\perp=0$ the k-dependence of the
dispersion relation yields
\begin{equation}
E_{\pm}(k)=\pm(1+2Q)|\sin k|.
\end{equation}
The ground state corresponds to the situation where the lower band is
fully occupied and the upper band is empty. A fermion $\alpha$
created in the upper band produces
the elementary excitation in the system and the corresponding energy
excitation is given only by the dispersion relation of the upper band, namely:
\begin{equation}
\epsilon(k)=(1+2Q)|\sin k|.
\end{equation}
One can calculate $Q$ and finds $1+2Q\approx1.63$. This result
compares well with $\pi/2\approx 1.57$ in the exact solution of
Eq.\ (\ref{spectre1}). The interesting feature is that one recovers
smoothly the 1D limit of the dispersion relation by taking $m=0$.
The energy gap $E_g(J_\perp=0)$ is then equal to zero. When $J_\perp$ is
nonzero, the gap has the form
\begin{eqnarray}
E_g(J_\perp)&&=E_+({\bf k}=(0,\pi))\cr
&&=\bigl(m^2(1+J_\perp/2)^2+(1+2P)^2
(J_\perp/2)^2\bigr)^{1/2}
\label{gap}
\end{eqnarray}
which reduces to
\begin{equation}
E_g(J_\perp)={{(1+2P)}\over2}J_\perp
\end{equation}
for $m=0$. Eq.\ (\ref{gap}) is obtained by calculating the difference
between the ground state energy
$$
E_{GS}=JQ^2+J_\perp P^2+(J+J_\perp){m^2\over4}-{1\over2}
\int {{{d^2k}\over{2\pi}}}E_+({\bf k})
$$
and the first excited state energy
$$
E_{EX}=JQ^2+J_\perp P^2+(J+J_\perp){m^2\over4}-{1\over2}
\int_ {{\bf k}\ne(0,\pi)}{{{d^2k}\over{2\pi}}}E_+({\bf k})
+{1\over2}E_+({\bf k}=(0,\pi)).
$$
For $m=0$ the results of the numerical calculation for the set of
Eqs.\ (\ref{moyen}) are displayed in Figs. 3 and 4. The
parameters $Q$ and $P$ show no simple dependence on $J_\perp$.
The energy gap, which is displayed in Fig. 4, has a linear behavior
in $J_\perp\gg 1$. This is in good qualitative agreement with
the simple limit $J_\perp=\infty$. It has a more complicated
dependence for intermediate transverse coupling because of the
$J_\perp$-dependence of $P$ (Fig. 3). For small $J_\perp$,
$E_g(J_\perp)$ has a simple power law form
\begin{equation}
E_g(J_\perp)\approx c{(J_\perp)}^g
\end{equation}
where the constant $c=0.76$ and the exponent $g=1.15$. This result compares
qualitatively well with that of Strong and Millis \cite{Strong} who
studied this problem in the case of z-anisotropy in the
parallel Heisenberg coupling.

	The N\'eel-flux-phase state has a nonzero $m$ for $J_\perp<1.76$.
The different parameters of this state are displayed in Fig. 5. The gap
is found to go to a finite limit when $J_\perp \to 0$. The finite
magnetization in this state implies broken rotational symmetry.
Gapless collective modes related to spin wave excitation would then
exist in the gap. The spin wave theory goes beyond the mean field
approximation. For nonzero $m$, the quantum fluctuations
due to spin wave excitations would
have drastic repercussions on the
value of $m$. As in 1D, these fluctuations destroy long range order.
Indeed, for the ladder, the corrections to the local magnetization
in the standard spin wave theory can be calculated and are found
to be logarithmically singular
\begin{equation}
\Delta \langle S^z\rangle\sim\int {dk\over k}\sim -\infty.
\end{equation}
This implies that spin wave theory is not self-consistent and no long range
order can occur at zero temperature.
So the flux-phase solution, $m=0$, is more adequate to describe
the AF correlations
even if its ground state energy is slightly higher than that of the
N\'eel-flux-phase state as shown in Fig. 6. All the physical information
the nonzero solution contains is that the AF correlations are more
important for $0<J_\perp<1.76$ as we see on Fig. 5 because the magnetization
is zero for $J_\perp>1.76$, but, to our opinion, these correlations
are not strong enough to induce
long range order, (imagine that we can solve exactly this problem,
then one would find a zero magnetization for any value of $J_\perp$).
The N\'eel-flux-phase and flux-phase states give
the same dispersion relation for $J_\perp>1.76$.
\section{numerical investigation with the DMRG}
We have used the recently introduced density matrix renormalization group
method \cite{White} to find the ground state energy of the Heisenberg ladder.
Our Fortran codes are written with
the version of the infinite lattice method with open boundary
conditions. We first find the ground state wavefunction of the finite
$2\times 7$ system and start the renormalization process by keeping
80 states in
each block. Unlike the one-dimensional version suggested by White \cite{White}
we insert only one pair of new sites in the middle of the two blocks to form
the new superblock in each
renormalization procedure so that the size of
the Hilbert space is kept within
our computer's capacity. We find this works reasonably well for the ground
state energy shown as crosses in Fig. 6.
For the singlet-triplet energy gap however, it works less well.
Nevertheless, the gap we obtain numerically shows convincingly that it vanishes
only as the $J_\perp$ goes to zero as should be apparent from the crosses of
Fig. 4. The numerical results will be given in detail in a forth coming work
\cite{Chen}
\section{Conclusion}
In conclusion, our mean-field theory (flux-phase) describes accurately
the low lying excitations of the AF Heisenberg model on the ladder. The
gap is found to increase smoothly with $J_\perp$. Its behavior as a
function of $J_\perp$ is shown in Fig. 4. The quantitative agreement between
the analytical approach and the DMRG numerical solution is very good.

	The accuracy of the flux-phase in the case of the ladder is a
precursor of high dimensionality physics since a finite flux can
exit only in dimensions
higher than 1. The spin energy excitations have a gap. Now, when
charge degrees of freedom are introduced, the situation becomes
more complicated.
However, if we assume that a finite doping, $\delta_c$, is required
to bring the
gap to zero, then one can conclude that the 1D Luttinger liquid state is
unstable for $\delta<\delta_c$ in the sense that the spin correlations
decrease algebraically in 1D rather than exponentially
for finite $J_\perp$. The
system does not belong to the same universality class when $J_\perp>0$ as
that of the 1D system.

\acknowledgements
We are grateful to A.-M. S. Tremblay for helpful and
interesting discussions. One of us (LC) would like to thank E. S{\o}rensen
for discussion on D.M.R.G. This work was supported by the Natural
Sciences and
Engineering Research Council of Canada (NSERC) and the Fonds pour
la formation de chercheurs et l'aide a la recherche from the Government of
Qu\'ebec (FCAR).
\begin{figure}
\caption{The ladder system.}
\end{figure}
\begin{figure}
\caption{The flux per plaquette is equal to $\pi$ on average}
\end{figure}
\begin{figure}
\caption{The parameters $Q$ and $P$ as a function of $J_\perp$ for
$m=0$}
\end{figure}
\begin{figure}
\caption{The energy gap as a function of $J_\perp$. The full and dashed
lines are respectively from our mean field treatment ($m=0$)
and the result of perturbation theory of Ref. 6.
The $+$'s are the result of the D.M.R.G
calculation.}
\end{figure}
\begin{figure}
\caption{The parameters of the nonzero magnetization solution of
the mean field approximation (plotted as a function of $J_\perp$)}
\end{figure}
\begin{figure}
\caption{The ground state energy as a function of $J_\perp$.
The full and dashed lines correspond respectively to zero and nonzero
magnetization. The $+$'s are the result of the D.M.R.G calculation.}
\end{figure}
\end{document}